\documentclass[]{spie}  

\usepackage{amsmath,amsfonts,amssymb}
\usepackage{graphicx}
\usepackage[colorlinks=true, allcolors=blue]{hyperref}

\title{High contrast imaging at the photon noise limit with self-calibrating WFS/C systems}

\author[a,b,c,d]{Olivier Guyon}
\author[e]{Barnaby Norris}
\author[e]{Marc-Antoine Martinod}
\author[a]{Kyohoon Ahn}
\author[e]{Peter Tuthill}
\author[b]{Jared R. Males}
\author[e]{Alison Wong}
\author[a,f,g]{Nour Skaf}
\author[a]{Thayne Currie}
\author[h]{Kelsey Miller}
\author[h]{Steven P. Bos}
\author[a]{Julien Lozi}
\author[a]{Vincent Deo}
\author[a,d,f]{S\'ebastien Vievard}
\author[i]{Ruslan Belikov}
\author[b]{Kyle van Gorkom}
\author[b]{Sebastiaan Y. Haffert}
\author[j]{Benjamin A. Mazin}
\author[k]{Michael Bottom}
\author[l]{Richard Frazin}
\author[b]{Alexander Rodack}
\author[m]{Tyler D. Groff}
\author[n]{Nemanja Jovanovic}
\author[o]{Frantz Martinache}

\affil[a]{Subaru Telescope, National Astronomical Observatory of Japan, National Institutes of Natural Sciences (NINS), 650 North A`oh\={o}k\={u} Place, Hilo, HI 96720, United States}
\affil[b]{Steward Observatory, University of Arizona, Tucson, AZ 87521, United States}
\affil[c]{College of Optical Sciences, University of Arizona, Tucson, AZ 87521, United States}
\affil[d]{Astrobiology Center of NINS, 2 Chome-21-1, Osawa, Mitaka, Tokyo, 181-8588, Japan}
\affil[e]{Sydney Institute for Astronomy, Institute for Photonics and Optical Science, School of Physics, University of Sydney, NSW 2006, Australia}
\affil[f]{Observatoire de Paris, LESIA, 5 Place Jules Janssen, 92190 Meudon, France}
\affil[g]{Department of Physics and Astronomy, University College London, London, United Kingdom}
\affil[h]{Leiden Observatory, Leiden University, Niels Bohrweg 2, 2333 CA, Leiden, The Netherlands}
\affil[i]{NASA Ames Research Center, Moffett Blvd, Mountain View, CA 94035, United States}
\affil[j]{University of California Santa Barbara, Santa Barbara, CA 93106, United States}
\affil[k]{Institute for Astronomy, University of Hawai'i, 640 North A`oh\={o}k\={u} Place, Hilo, HI 96720, United States}
\affil[l]{Department of Climate and Space Sciences and Engineering, University of Michigan, Ann Arbor, MI 48109, United States}
\affil[m]{Goddard Space Flight Center, 8800 Greenbelt Rd, Greenbelt, MD 20771, United States}
\affil[n]{California Institute of Technology, 1200 E California Blvd, Pasadena, CA 91125, United States}
\affil[o]{Laboratoire Lagrange, Universit\'e C\^{o}te d'Azur, Observatoire de la C\^{o}te d'Azur, CNRS, Parc Valrose, B\^{a}t. H. FIZEAU, 06108 Nice, France}

\authorinfo{Further author information: (Send correspondence to O.G.)\\O.G.: E-mail: guyon@naoj.org}

\begin{document} 
\maketitle

\begin{abstract}
High contrast imaging (HCI) systems rely on active wavefront control (WFC) to deliver deep raw contrast in the focal plane, and on calibration techniques to further enhance contrast by identifying planet light within the residual speckle halo. 
Both functions can be combined in an HCI system and we discuss a path toward designing HCI systems capable of calibrating residual starlight at the fundamental contrast limit imposed by photon noise.
We highlight the value of deploying multiple high-efficiency wavefront sensors (WFSs) covering a wide spectral range and spanning multiple optical locations.
We show how their combined information can be leveraged to simultaneously improve WFS sensitivity and residual starlight calibration, ideally making it impossible for an image plane speckle to hide from WFS telemetry. 
We demonstrate residual starlight calibration in the laboratory and on-sky, using  both a coronagraphic setup, and a nulling spectro-interferometer. In both case, we show that bright starlight can calibrate residual starlight.
\end{abstract}


\section{Scientific motivation}
\label{sec:fundlimits}

Direct imaging of exoplanets and circumstellar disks is key their detailed characterization, but is challenging due to the small angular separation between objects of interets and the much brighter host stars. 

\begin{figure}[h]
    \centering
    \includegraphics[width=17cm]{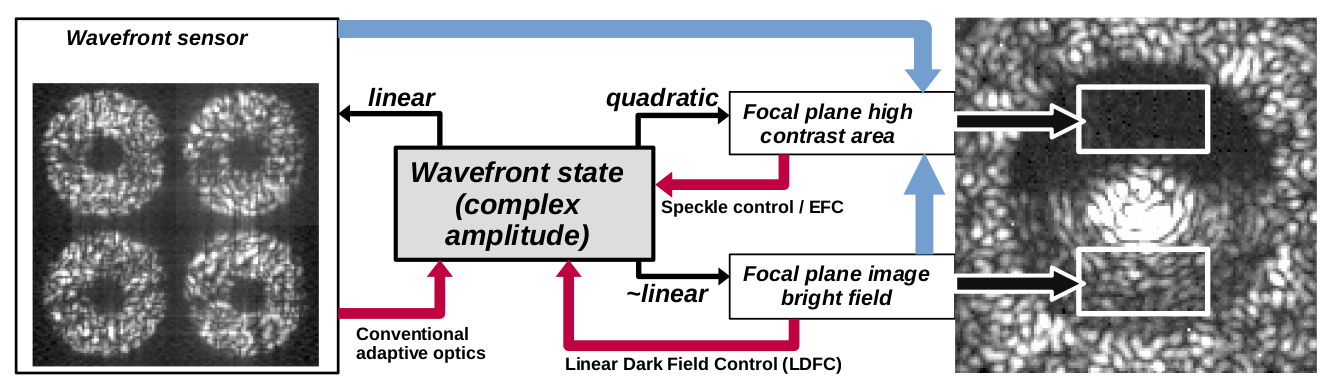}
    \vspace*{0.3cm}
    \caption{A high contrast imaging systems may include wavefront sensor(s) and focal plane image(s). Data obtained by camera(s) is related to the input wavefront state, which is not directly measured. Wavefront control loops (red arrows) use measurements to compensate for residual wavefront errors and deliver/maintain a high contrast area in the science focal plane (top right). Additionally, starlight measurements may be used to calibrate residual light in the high contrast area, as shown with the blue arrows.}
    \label{fig:HCIsystemarch}
\end{figure}

A high contrast imaging (HCI) system, as depicted in Figure \ref{fig:HCIsystemarch}, may include dedicated wavefront sensor(s) and focal plane image(s). By convention, a wavefront sensor is defined here as an optical device optimized for wavefront measurement, providing, in the small wavefront aberration regime, a linear relationship between input wavefront state and sensor signal.

Most ground-based HCI systems rely on pupil-plane wavefront sensor(s) to measure the fast, $\mu$m-scale wavefront aberrations induced by atmospheric turbulence. Pupil plane sensors can provide a linear signal over a wide range of wavefront aberration, making them ideal for high speed wavefront control. 

HCI systems envisioned for space missions, on the other hand, rely on focal plane image(s) to measure residual aberrations. This choice is motivated by the very small level or allowable residual wavefront error required to meet the $10^{10}$ contrast level for direct imaging of reflected light exoplanets. Such systems must overcome the non-linear relationship between input wavefront and measured focal plane intensity in the high contrast area, so wavefront modulation is employed to unambiguously measure focal plane complex amplitude in order to drive a wavefront control loop. Both approaches could fundamentally be combined as they are complementary - this is especially relevant for ground-based systems, where Extreme-AO level correction can deliver a stable focal plane suitable for focal plane wavefront control. Figure \ref{fig:HCIsystemarch} shows possible wavefront control strategies, each with a thick red arrow connecting the input sensor to the wavefront state upon which the control loop is acting.

Additionally, input sensors may also be used to estimate the residual starlight in the high contrast area, as shown by the blue arrows in Fig. \ref{fig:HCIsystemarch}. This can be performed as a post-processing operation, reconstructing and subtracting the unwanted residual starlight from science data. We discuss and explore this approach in this paper.

\subsection{Representative examples}

We consider a examples representative of challenging high contrast imaging observations with ground and space telescopes:
\begin{itemize}
    \item Earth-Sun system at 8pc distance, observed by a 4-m space telescope
    \item Earth-size planet orbiting in the habitable zone of a M4 type star at a 4pc distance, observed by a 30-m ground-based telescope
\end{itemize}

\begin{table}[ht]
\caption{Observation examples} 
\label{tab:HCIobs}
\begin{center}       
\begin{tabular}{|l|c|c|} 
\hline
            & Space-4m-Earth-G2 &  Ground-30m-Earth-M4\\
\hline
\hline
Star &  G2 at 8pc &  M4 at 4pc \\
\hline
Bolometric luminosity [$L_{Sun}$] &  1.000 & 0.0072  \\
\hline
Planet orbital radius [au] &  1.0 &  0.085 \\
\hline
Maximum angular separation [arcsec] &  0.125 & 0.021 \\
\hline
Reflected light planet/star contrast &  1.5e-10 & 2.1e-8 \\
\hline
\hline
Telescope diameter [m] & 4 & 30 \\
\hline
Science spectral bandwidth &  20\% &  20\%  \\
\hline
Central Wavelength &  797 nm (I band) &  1630 nm (H band) \\
\hline
Maximum angular separation [$\lambda$/D] & 3.0 & 1.9 \\
\hline
Efficiency &  20 \% &  20 \%  \\
\hline
Total Exposure time &  10 ksec &  10 ksec  \\
\hline
\hline
Star brightness & $m_I = 3.60$ & $m_H = 5.65$ \\
\hline
Photon flux in science band (star) & 7.37e8 ph/s & 5.62e9 ph/s \\
\hline
Photon flux in science band (planet) & 0.11 ph/s & 118 ph/s \\
\hline
Background surf. brightness [contrast] & 3.5e-10 (zodi+exozodi) & 1e-5 (starlight)\\
\hline
Background flux in science band & 0.26 ph/s & 56200 ph/s\\
\hline
\hline
{\bf Photon-noise limited SNR (10 ksec)} & 18.1 & 49.7 \\
\hline

{\bf Post-processing timescale (SNR=10 at planet flux)} & 50 mn & 7 mn\\
\hline
{\bf WFS timescale (SNR=10 at background flux)} & 6 mn & 1.8 ms\\
\hline
\end{tabular}
\end{center}
\end{table} 

Table \ref{tab:HCIobs} lists key parameters relevant to the photometric detection of the planet for each case.

Observations are background-limited in both cases. For the space-based observation, we adopt a $m_V = 21\:\mathrm{arcsec}^{-2}$ combined zodiacal + exozodiacal background surface brightness, with $V-I=0.7$ matching the stellar spectrum, resulting in a background is 2.4 $\times$ brighter than the planet. The planet photon rate, at 0.11 ph/s, requires a 50mn integration to reach SNR=10. Thanks to the larger collecting area, the planet photon rate is much larger ($>$100 ph/s) for the ground-based observation, but the background is also significantly higher.  

The bottom part of the table provides timescales relevant to the identification of speckles in the image. The {\bf time to SNR=10 on planet flux} indicates the exposure time required to detect in the image speckles at a level comparable to the planet image. In the absence of independent speckle calibration, the speckle field should remain stable over multiple such timescales. The last row shows the {\bf time to SNR=10 on speckle at background}, indicating how much time is needed for the instrument to measure speckles at the raw contrast level. Without a separate wavefront sensing scheme, the speckle field must be stable at the raw contrast level over this timescale.

In both cases, the postprocessing and WFS timescales are long compared to possible sources of wavefront disturbances, driving optical stability requirements to become very challenging to meet. In deriving these quantities, we however only considered light available in the high-contrast area (dark hole) of the science image. A more efficient approach would be to use most of the available starlight for wavefront sensing and calibration. This is the goal of our study.

\section{Focal plane speckle calibration}
\label{sec:specklecalib}

\subsection{Goals}

In this first example, we consider dark hole (DH) calibration in a focal plane image where the DH occupies half of the field, while the other half is significantly brighter, and referred to as the bright field (BF). The configuration is shown in Fig. \ref{fig:HCIsystemarch} on the rightmost image. In this configuration, our goal is to have the BF be used to calibrate the DH.

The approach is related to linear dark field control (LDFC) \cite{2021A&A...646A.145M}, where a linear control loop uses the BF as input for wavefront control. LDFC has been demonstrate to stabilize the DH both in the laboratory \cite{2020PASP..132j4502C} and on-sky \cite{2021arXiv210606286B}. Here, we explore extending LDFC in a nonlinear DH calibration algorithm. Ideally, an algorithm would have as input the BF and return an estimate of the DH. This estimate is computed for each exposure and subtracted from the measured DH to remove speckles due to time-variable wavefront errors.

Important for this approach is that the BF is engineered to contain all necessary wavefront information to calibrate the DH. 
That is, the BF should exhibit a unique response to sign changes of even and odd wavefront modes, which is not necessarily always the true for every system. 
For example, focal-plane wavefront sensing with the BF of a vAPP coronagraph is enabled by a pupil amplitude asymmetry in the coronagraph's design\cite{bos2019focal}. 
This allowed for the successful LDFC experiments presented in Ref. \citenum{2021A&A...646A.145M, 2021arXiv210606286B}.
When amplitude aberrations need to be calibrated as well, for example to deal with the asymmetric wind driven halo\cite{cantalloube2018origin} (interference of AO-lag error and scintillation), then the BF also needs to be designed to have a unique response to these aberrations as well\cite{bos2020sky}.   

\subsection{Experimental setup}

Data was acquired on the Subaru Coronagraphic Extreme Adaptive Optics (SCExAO)\cite{jovanovic2015subaru, Lozi2018SCExAO} instrument using its internal light source with a Lyot-type coronagraph. The deformable mirror (DM) was configured to yield a $\approx$ 1e6 contrast dark hole in H-band. 128x128 pixel images were acquired at 1550nm (25nm bandpass) at a 7 kHz framerate using a C-RED one camera with a SAPHIRA-type imaging array. During the acquisition, simulated dynamical wavefront aberrations were added to the system deformable mirror to produce time-variable speckles in the DH. Corresponding BF modulations were recorded in the images.

\begin{figure}[h]
    \centering
    \includegraphics[width=17cm]{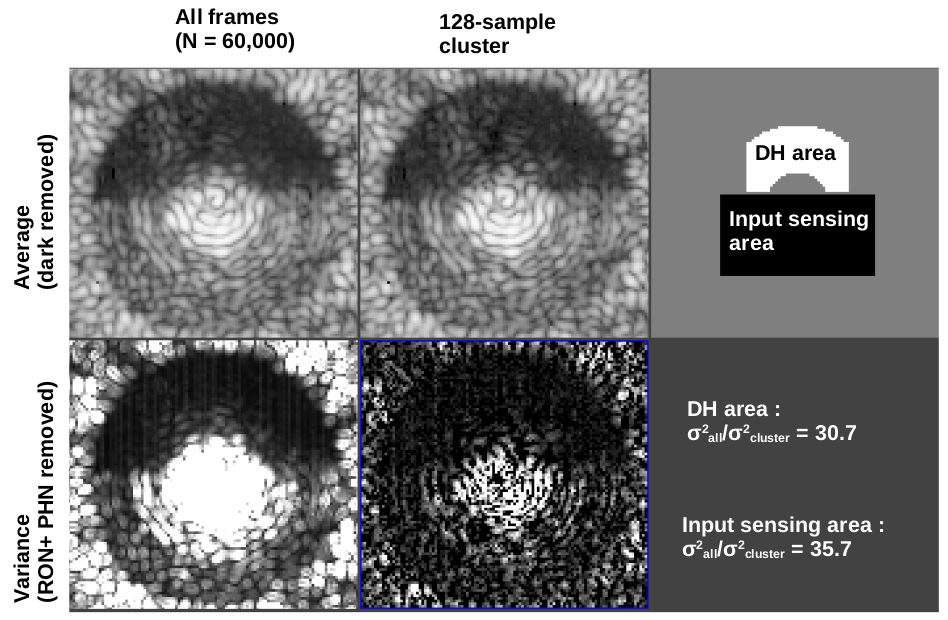}
    \caption{Dark hole calibration using high contrast image bright field.}
    \label{fig:DHcalibSCExAO}
\end{figure}

\subsection{Algorithm}

The goal of the experiment is here to quantify how well the image BF can constrain the DF. The approach we adopt is to identify within a large set of images a subset of images with nearly identical BF realisations, and measure the statistical properties of the corresponding DH images. If the DH can be reconstructed from the BF, then images for which the BF is nearly identical should also have nearly identical DHs. If, however, several DH solutions map to the same BF, then our analysis would reveal a large scatter in DH realisations within the BF-selected set.

This statistical approach does not require an algorithm to be developed to compute DH from BF, and does not make any assumption of what such an algorithm is. If identical BFs correspond to identical DHs, then a DH reconstruction algorithm does exist. While it may be conceivable to construct a correspondence table between the two, this may not be practical in high dimension space, and more efficient techniques such as neural networks may be employed, as illustrated in \S \ref{sec:VAMPIRES}.

The steps of the algorithm are:
\begin{itemize}
    \item Identify clusters of images with similar BF realisations
    \item Select optimal cluster, according to cluster sample size, cluster diameter, and possibly DH flux (see \S \ref{sec:GLINT})
    \item Measure variance in DH intensity map within cluster, and compare to full input sample
\end{itemize}

\subsection{Results}

Results are compiled in Figure \ref{fig:DHcalibSCExAO}. The average of all 60000 frames (top left) shows the BF in the lower half of the image and the DH in the top half. The intensity variance across the 60000 images is shown for each pixel of the image at the bottom left, with both readout noise and photon noise variance terms subtracted to reveal actual speckle intensity variance.

The area used to select clusters of BF-similar images is shown in black in the top right image: it is the set of pixels used to compute a distance metric between BF realisations. A cluster of 128 images with similar BFs was identified, and the corresponding average and variance images shown at the center column with identical brightness scales to the ensemble average and variance on the left.

To measure the algorithm performance, the variance across a set of images is computed. Noting $x,y$ the image spatial coordinates and $k$ the image index, the BF variance across the whole set of images is :
\begin{equation}
    \sigma^2_{BF,all} = \frac{1}{N_{all} \: N_{pixBF}} \sum_{k,(x,y) \subset BF} \left( I_k(x,y,k) - \left( \frac{1}{N_{all}} \sum_{k1} I(x,y,k1) \right) \right) ^2
\end{equation}
where $I(x,y,k)$ is the intensity of pixel with spatial coordinates $(x,y)$ in frame number $k$, $N_{all}$ is the total number of frames, and $N_{pixBF}$ is the number of pixels $(x,y) \subset BF$ in the bright field. Similarly, variance $\sigma^2_{DH,all}$ is defined for the DH, and $\sigma^2_{BF,cluster}$ is the variance across the set of images within the selected cluster. The DH geometry used for the variance computation is shown if Fig. \ref{fig:DHcalibSCExAO} top right panel.

Figure \ref{fig:DHcalibSCExAO} shows that the variance within the BF is 35.7 $\times$ smaller within the selected 128-sample cluster than across the full dataset. This is to be expected, as the cluster selection is based on minimizing the BF variance across the selected set of images. The corresponding measured DH variance is 30.7 $\times$ smaller within the cluster set than across the full input dataset. This last result demonstrates that images with similar BFs also have similar DHs. {\bf Image selection using BF intensity successfully constrains DH intensity, demonstrating that a BF-to-DH calibration algorithm can be derived to calibrate residual speckles in high contrast images}.

\begin{figure}[h]
    \centering
    \includegraphics[width=16cm]{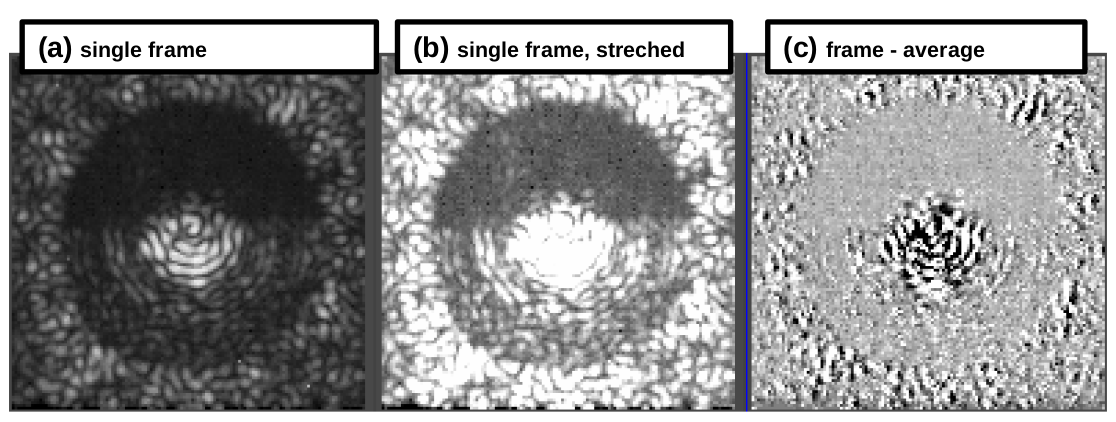}
    \vspace*{0.3cm}
    \caption{Single frame within the 128-sample cluster. The frame is shown with two different brightness scales (a and b) to highlight the BF and DH image areas. The difference between the frame and the average over the full dataset (c) shows strong signal within the BF, but no detectable signal in the DH due to low SNR.}
    \label{fig:HCIsingleframe}
\end{figure}

A single frame within the selected set is shown in Fig. \ref{fig:HCIsingleframe}, along with its deviation from the average intensity image across the full set of images. The BF signal used for the selection is strongly visible in image (c), while the DH area is indistinguishable from the average DH intensity due to low SNR. The selection we performed would therefore not have been possible from the DH intensity alone. This also demonstrates the technique's ability to reconstruct the DH speckle map to an accuracy better than the readout and photon noise. 

This last point is essential to the approach's main goal: wavefront variations can be tracked and their effect on the DH calibrated faster than the WFS timescale listed in Table \ref{tab:HCIobs}. For example, a mechanical vibration may create a time-variable speckle that would be indistinguishable from a planet within the DH. The same vibration would modulate the BF intensity, and a BF-to-DH algorithm would detect this modulation and calibrate out the offending DH speckle thanks to higher BF SNR. While the LDFC algorithm leverages the same SNR gain, it would require the LDFC control bandwidth to exceed the vibration timescale. This is not required for postprocessing, as variance analysis of the BF would reveal the vibration, provided that the image framerate (for the BF) is sufficiently fast to resolve the vibration.

\section{PSF reconstruction from WFS telemetry}
\label{sec:VAMPIRES}

While \S \ref{sec:specklecalib} demonstrates that a unique mapping exists between BF and DH, we have not constructed an algorithm to transform BF measurements into DH estimates. The non-linear relationship between the two spaces eliminates conventional linear reconstructors, and the high dimension of the input BF image makes a lookup table impractical. We demonstrate in this section that a neural network is a viable approach to address this challenge.

\label{sec:WFStoPSF}
\begin{figure}[h]
    \centering
    \includegraphics[width=14cm]{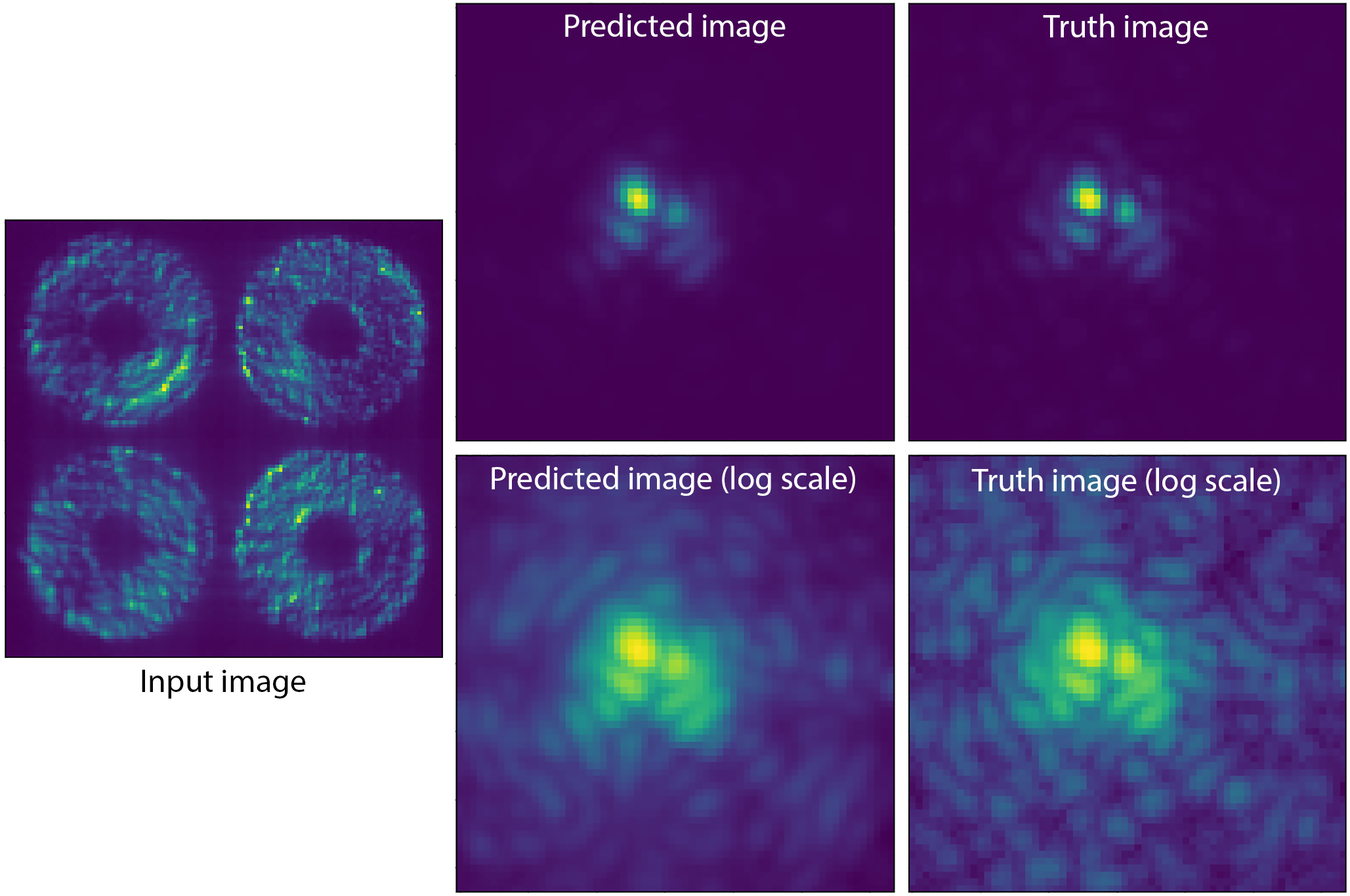}
    \vspace*{0.3cm}
    \caption{Prediction of the PSF from the Pyramid WFS data for an on-sky observation, using a neural network. The predicted PSF image (centre) is determined entirely from the current WFS image (left), and is seen to closely match the true PSF measured at that instant (right column). This example shows a PSF with a large amount of wavefront error (including strong coma) to provide a clear illustration.}
    \label{fig:wfspsf}
\end{figure}

Wavefront sensors are commonly used to drive an adaptive optics control loop. It is also possible to reconstruct the instantaneous PSF from the current wavefront sensor data. To utilise the greatest possible amount of wavefront sensor information, it is desirable for the reconstruction to be performed using the raw wavefront sensor image, rather than the modal basis used in the AO system. However the relationship between the pixel intensities in the wavefront sensor image and the pixel intensities in the focal plane is non-linear, precluding the use of a simple linear reconstruction. Instead, a neural network is used, as these are highly capable at non-linear inference tasks. 

Results of a PSF reconstruction for on-sky data are shown in Figure \ref{fig:wfspsf}. Here, a fully-connected neural network consisting of two 2000-unit layers was trained on 5 minutes of on-sky data, consisting of synchronised images from the pyramid wavefront sensor camera and VAMPIRES visible camera\cite{2015MNRAS.447.2894N} (wavelength 750~nm), running at approximately 500 frames/sec. The network used ReLU activation functions and dropout between each layer as a regularizer, the latter proving to be crucial for successful reconstruction. While the fully-connected network shown here provides good results, certain advantages (such as reduced parameter number and resistance to pupil alignment drift) could be expected from a convolutional neural network, which is the focus of a current study.

\section{Photonic nulling calibration}
\label{sec:GLINT}

While section \ref{sec:specklecalib} demonstrates that a mapping from BF to DF exists, we provided no practical solution to construct a BF-to-DH calibration algorithm. Due to the large number of input dimensions, a brute-force approach is not feasible. The BF-to-DH relationship may also slowly evolve due to slight changes in optical alignment or optical surfaces figures within the coronagraph optical train, rendering old BF-to-DH calibrations stale.

\begin{figure}[h]
    \centering
    \includegraphics[width=16cm]{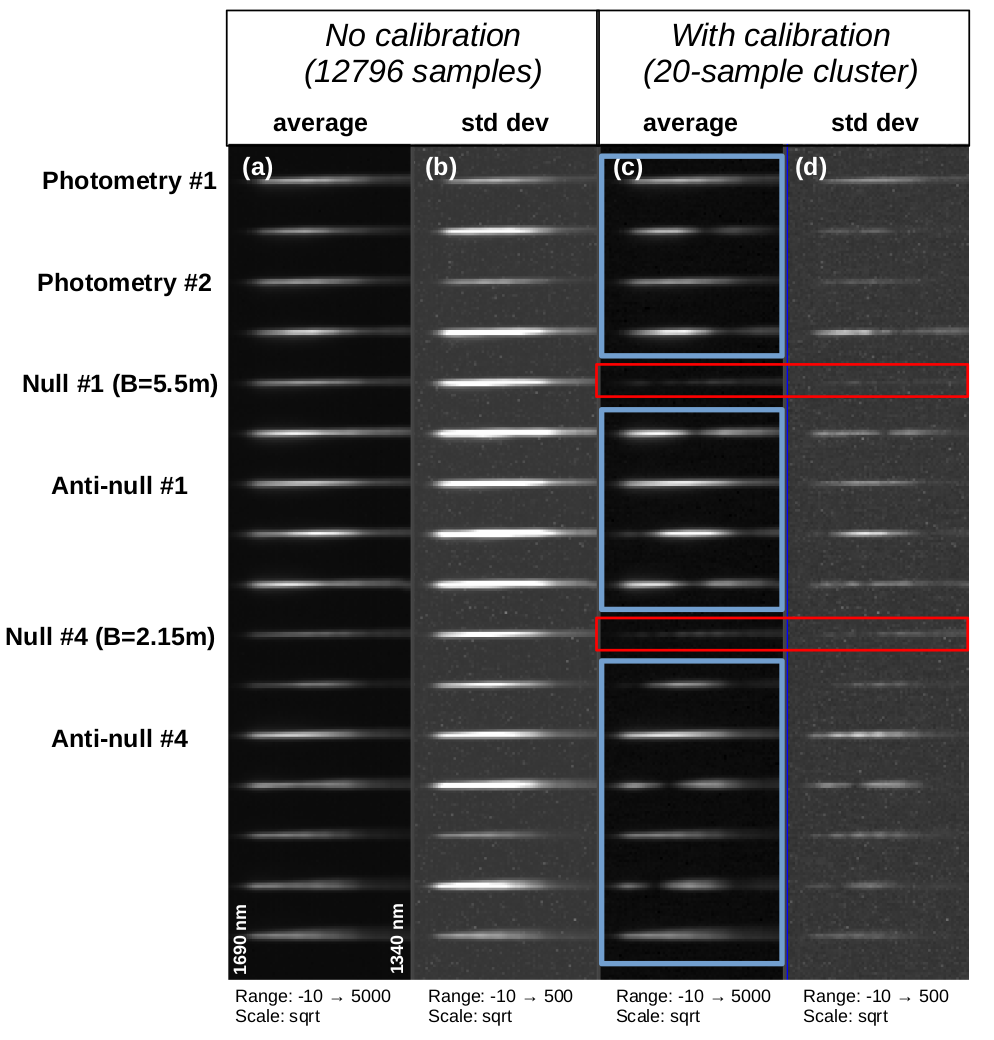}
    \vspace*{0.3cm}
    \caption{Null calibration with the GLINT photonic nuller: Laboratory demonstration.}
    \label{fig:GLINTlab}
\end{figure}

A photonic nuller is an alternative solution to the high contrast imaging challenge. Unlike a coronagraph constructed from bulk optics between which light freely propagates, the photonic nuller couples starlight into a small number of coherent singlemode waveguides. The waveguides are coherently combined to produce starlight destructive interference in null output(s). Bright starlight is directed to bright outputs which measure the intensity in input waveguides (photometry output(s)) and phase offset between input waveguides (WFS output(s)). The photonic nuller concept and its implementation are discussed in publications from the GLINT instrument team\cite{2020MNRAS.491.4180N, Martinod2021NatCo}.

The photonic nuller approach to high contrast imaging appears to be largely immune from the challenges affecting coronagraph systems with bulk optics:
\begin{itemize}
    \item Starlight is coupled in a small number of coherent waveguides. At each wavelength, light into the photonic device input is fully described by phase and amplitude (and possibly polarization), so the number of dimension in the input is a few times the number of waveguides
    \item The relationship between input variables (phase and amplitude of each waveguide) and output intensities is entirely established within the photonic chip so it is significantly more stable than an optical train of optical components subject to relative misalignments.
\end{itemize}

We tested the DH calibration approach on the GLINT instrument\cite{Martinod2021NatCo} installed on the Subaru Telescope. Figure \ref{fig:GLINTlab} shows results obtained with the internal light source and dynamical wavefront aberrations injected using SCExAO's deformable mirror. The leftmost image (a) is the average GLINT detector image across during the experiment. GLINT's 16 output waveguides are wavelength-dispersed and re-imaged on a nearIR camera, so the image shows 16 spectra, with wavelength ranging from 1340 nm (right edge) to 1690 nm (left edge). Images are acquired at 1.4 kHz framerate to freeze atmospheric turbulence and vibrations. Two null output channels are labelled in the figure, and should be dark in the absence of input wavefront errors. Anti-null outputs are where fully constructive interference should occur for a flat wavefront. Photometry channels track the amount of flux in input waveguides. Other input encode differential phase between input channels as interference fringes. Null \#1 is the destructive interference between two widely separated (B $=$ 5.5 m center-to-center) apertures on the 8.2 m Subaru telescope pupil, while null \#4 corresponds to a shorter (B $=$ 2.15 m center-to-center) baseline.

The left half of Fig. \ref{fig:GLINTlab} shows the average (a) and standard deviation (b) of all frames. Due to varying wavefront errors, null outputs are relatively bright, especially null \#1 which is highly sensitive to wavefront tip-tilt. Since wavefront variations in this test are larger than 1 radian, the average traces are mostly devoid of fringes, and standard deviation is comparable to average intensity. We note that the standard deviation is somewhat smaller, but still noticeable, for the photometric outputs, as the input wavefront errors is large enough to induce tip-tilt across input subapertures, resulting in a time-variable coupling efficiency loss.

We defined the BF selection area as all non-null output channels, and DH as the two null outputs. The BF used for frame selection is outlined by the blue rectangles in Fig. \ref{fig:GLINTlab} panel (c). The ouput DH consisting of two null channels is shown by the red rectangles. 

A 20-sample cluster is selected based on BF similarity (blue rectangles), with an added constraint on total average flux within the null output channels. The corresponding average (c) and standard deviation (d) images demonstrate, as in \S \ref{sec:specklecalib}, improved stability of both BF and DH. In this experiment, the input WF errors were considerably larger so the interferometric signal (fringes) was washed out in the whole set average, and the average null depth was poor. The average of the BF selection shows a clear WFS signal and maintains good starlight suppression in the null outputs. 

\begin{figure}[h]
    \centering
    \includegraphics[width=16cm]{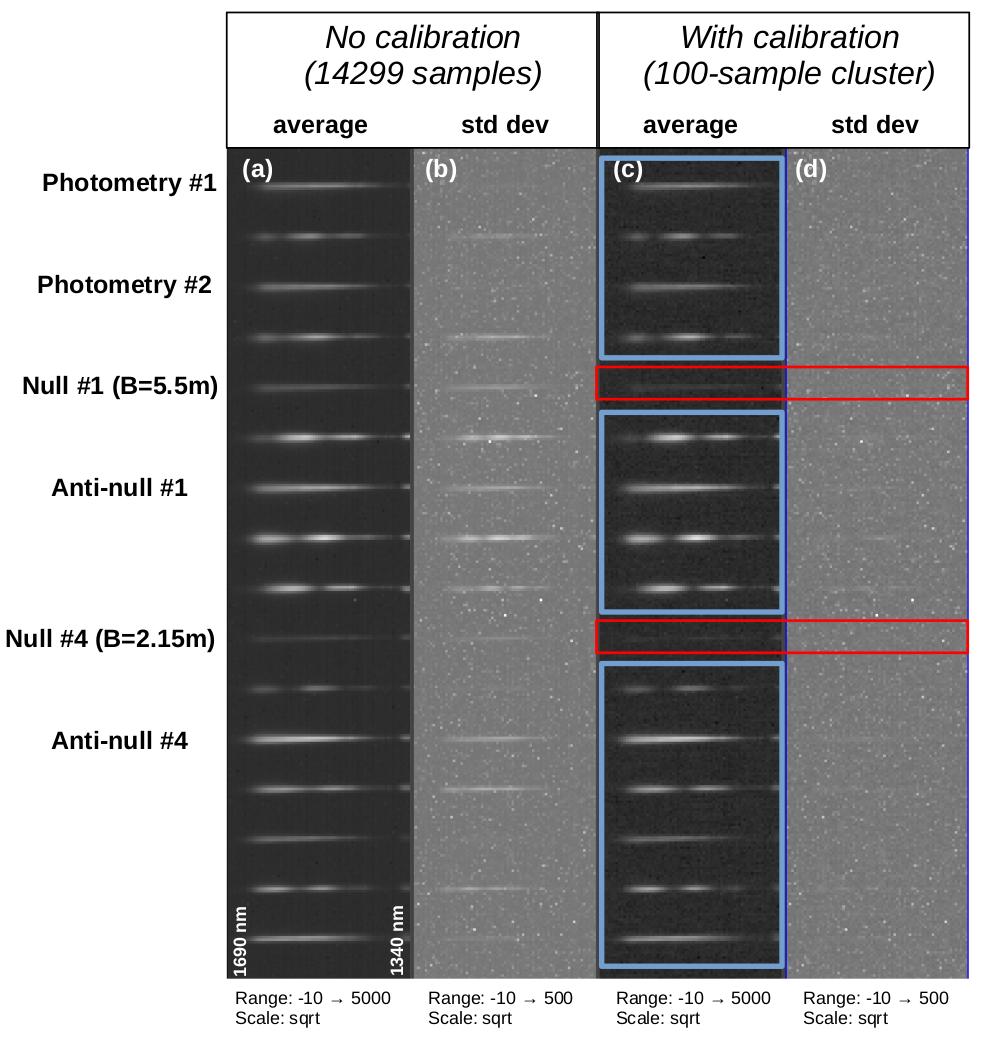}
    \vspace*{0.3cm}
    \caption{Null calibration with the GLINT photonic nuller: on-sky demonstration.}
    \label{fig:GLINTsky}
\end{figure}

We also performed the experiment on-sky as shown in Fig. \ref{fig:GLINTsky}. The star $\alpha$ Boo (Arcturus) was observed with the same setup. Input wavefront errors are smaller than in the laboratory demonstration, so fringes are visible in the average image (a). We choose here a larger 100-sample cluster to mitigate readout noise and photon noise. The null is deeper and stable within the cluster, demonstrating the BF selection does successfully identify high-quality frames within the dataset. The standard deviation in the null channels is small enough that it could not be measured, as it is well below the detector noise level. {\bf The quality (null depth and stability) of the BF-selected dataset is therefore significantly better than possible with a DH-selected dataset}. GLINT's detection of $\alpha$ Boo's finite angular diameter is visible as a stable non-null flux in null \#1.

On-sky results demonstrate uniqueness of null solution for a given BF measurement, which is the necessary condition for BF-to-DH algorithm. There is no evidence for a measurement null space which would induce a variation in the null ouputs without a corresponding signature in the bright channels.

\section{Conclusion}

We have demonstrated that bright starlight in wavefront sensor(s) and bright parts of images can be used to reliably calibrate residual starlight in the dark hole regions of high contrast images. Our preliminary lab and on-sky tests show that this DH estimation is more precise than the DH photon noise, so it may be possible to have self-calibrating HCI systems operate at the photon noise limit imposed by total surface brightness.

We have shown that there exists an unambiguous BF-to-DH mapping, and in our tests, we have shown that there is no uncalibrated DH variations. While our tests are encouraging, we have not yet developed a reliable, practical BF-to-DH reconstruction algorithm, but show that a neural network can solve a closely related PSF reconstruction problem. We note that our lab and on-sky tests were of short duration, and that the underlying BF-to-DH relationship may change over longer period of time, possibly requiring frequent or continuous re-calibration. The approeach appears to be especially powerful for photonic nulling devices, where the starlight suppression and wavefront sensing functions are integrated in a single stable small-size device.

Our findings indicate that future HCI systems should run concurrently with multiple wavefront sensors (WFSs), so that the aggregate WFS information can be collected using as much starlight as possible to provide a high fidelity, high frame rate estimate of the DH speckle field. In such a system, science images can be calibrated to high accuracy, reducing chances for false positives and increasing data quality.

\acknowledgments

This work was supported by NASA grants \#80NSSC19K0336 and \#80NSSC19K0121. This work is based on data collected at Subaru Telescope, which is operated by the National Astronomical Observatory of Japan. The authors wish to recognize and acknowledge the very significant cultural role and reverence that the summit of Maunakea has always had within the Hawaiian community. We are most fortunate to have the opportunity to conduct observations from this mountain. The authors also wish to acknowledge the critical importance of the current and recent Subaru Observatory daycrew, technicians, telescope operators, computer support, and office staff employees.  Their expertise, ingenuity, and dedication is indispensable to the continued successful operation of these observatories. The development of SCExAO was supported by the Japan Society for the Promotion of Science (Grant-in-Aid for Research \#23340051, \#26220704, \#23103002, \#19H00703 \& \#19H00695), the Astrobiology Center of the National Institutes of Natural Sciences, Japan, the Mt Cuba Foundation and the director's contingency fund at Subaru Telescope. KA acknowledges support from the Heising-Simons foundation.

\bibliography{report} 
\bibliographystyle{spiebib} 

\end{document}